# Magnetoelectric effects of nanoparticulate Pb(Zr$_{0.52}$Ti$_{0.48}$)O$_3$–NiFe$_2$O$_4$ composite films

**Hyejin Ryu, P. Murugavel, J. H. Lee, S. C. Chae, and T. W. Noh[a]**

*ReCOE & FPRD, School of Physics and Astronomy, Seoul National University, Seoul 151-747, Korea*

**Yoon Seok Oh, Hyung Jin Kim, and Kee Hoon Kim**

*XMPL & FPRD, School of Physics and Astronomy, Seoul National University, Seoul 151-747, Korea*

**Jae Hyuck Jang and Miyoung Kim**

*School of Materials Science and Engineering, Seoul National University, Seoul 151-747, Korea*

**C. Bae and J.-G. Park**

*Department of Physics, Sungkyunkwan University, Suwon 440-746, Korea*

**Abstract**

We fabricated Pb(Zr$_{0.52}$Ti$_{0.48}$)O$_3$–NiFe$_2$O$_4$ composite films consisting of randomly dispersed NiFe$_2$O$_4$ nanoparticles in the Pb(Zr$_{0.52}$Ti$_{0.48}$)O$_3$ matrix. The structural analysis revealed that the crystal axes of the NiFe$_2$O$_4$ nanoparticles are aligned with those of the ferroelectric matrix. The composite has good ferroelectric and magnetic properties. We measured the transverse and longitudinal components of the magnetoelectric voltage coefficient, which supports the postulate that the magnetoelectric effect comes from direct stress coupling between magnetostrictive NiFe$_2$O$_4$ and piezoelectric Pb(Zr$_{0.52}$Ti$_{0.48}$)O$_3$ grains.

[a] Corresponding author: twnoh@snu.ac.kr

Magnetoelectric (ME) materials have attracted much research interest due to the coupling among the electric, magnetic, and elastic order parameters.[1–4] The figure of merit for ME materials is the magnetoelectric voltage coefficient ($\alpha_E$), which represents the amount of polarization induced under a given applied magnetic field.[5] Many workers have tried to combine perovskite ferroelectrics (such as $BaTiO_3$ and $Pb(Zr, Ti)O_3$) with spinel ferromagnetic materials (such as $CoFe_2O_4$ and $NiFe_2O_4$) to improve the $\alpha_E$ value in numerous bulk composites.[6–9] Recently, Zheng *et al.* reported a thin film type of ME material, composed of self-assembled $CoFe_2O_4$ spinel nanopillars in a $BaTiO_3$ perovskite matrix on a single crystal substrate.[10] In this intriguing composite thin film, nanopillars and matrix materials were grown with a high degree of crystallographic orientation. Good connectivity among the constituents might result in significant coupling between the piezoelectric and magnetostrictive phases to enhance the ME effects. However, no direct evidence of such ME effects has been reported yet.[10,11]

In this letter, we report another type of thin film composite ME material, in which ferromagnetic nanoparticles are dispersed with their crystal orientations aligned in the ferroelectric matrix. Since the ferromagnetic $NiFe_2O_4$ (NFO) phase exists as nanoparticles, instead of nanopillars, its microstructure can reduce possible leakage current paths through the ferromagnetic phase substantially. As a model system, we selected the $0.65Pb(Zr_{0.52}Ti_{0.48})O_3$ (PZT)–$0.35NiFe_2O_4$ (NFO) system due to its high piezoelectric and magnetostrictive properties.[9]

We fabricated the PZT–NFO films on (001) oriented 0.5% Nb-doped $SrTiO_3$ (Nb:STO) substrates using pulsed laser deposition. We used a KrF excimer laser (248 nm, Lambda Physik), with laser fluence and repletion rates of 1 J/cm$^2$ and 3 Hz, respectively. We deposited the films at 600°C at an oxygen partial pressure of 100 mTorr. To obtain the 0.65PZT–0.35NFO composite films, we used a modified target, joined by two sintered semicircular disks of PZT and NFO. We carefully chose the disk sizes and target rotations, and confirmed the composition of our deposited PZT–NFO films using electron probe microscope analysis.

The constituents of the PZT–NFO composite film have their crystal orientations aligned with respect to the substrate. Figure 1(a) shows the x-ray diffraction (XRD) $\theta$-$2\theta$ pattern around the (002) Nb:STO peak, measured using the synchrotron radiation source at the Pohang Light Source, Korea. The distinct set of PZT and NFO peaks confirms that the film contained coexisting PZT and NFO phases with no impurity phases. In addition, the perovskite PZT (002) and spinel NFO (004) peaks indicate that their $c$-axes should be aligned perpendicular to the substrate. Figure 1(b) displays the $\phi$ scan data recorded around the NFO (404), PZT (202), and Nb:STO (202) reflections, which show a set of four distinct peaks with 90° of separation. These XRD data confirm that the NFO and PZT grains are aligned in a specific orientation, such as (001)NFO//(001)PZT//(001)Nb:STO and [100]NFO//[100]PZT//[100]Nb:STO.

To obtain further microstructural information, we performed transmission electron microscope (TEM) studies with a Tecnai 20F field emission microscope. Figure 2(a) shows a low-resolution TEM image, in which the marked regions indicate that the NFO nanoparticles are well dispersed in the PZT matrix. We also obtained high-resolution images near the NFO–PZT boundary, as shown in Fig. 2(b) and (c). Figure 2(c) shows a well-defined boundary with clear atomic arrangements of the PZT matrix and NFO nanoparticles. We could index the spots in the selected area electron diffraction patterns, shown in Fig. 2(d), to PZT (220) and NFO (440). Both Fig. 2(c) and (d) indicate that the particles in the composite film are aligned with the [100] NFO // [100] PZT relationship. These TEM studies confirm that the NFO nanoparticles are dispersed randomly in the PZT matrix, as shown schematically in Fig. 2(e). This microstructure differs from that of the self-assembled nanopillars,[10] shown schematically in Fig. 2(f). The microstructure of our PZT–NFO composite film could significantly reduce leakage current paths in the ferroelectric and ME measurements.

Our PZT–NFO composite film demonstrated good ferroelectric and ferromagnetic properties. We measured the polarization *vs*. electric field (*P–E*) curves at various applied electric fields using a Precision workstation (Radiant Technologies). As shown in Fig. 3(a), the measured *P–E* loops had good ferroelectric characteristics: the remnant polarization, saturation polarization, and coercive field at the applied electric field of 800 kV/cm were 45 $\mu C/cm^2$, 60 $\mu C/cm^2$, and 275 kV/cm, respectively. These values are comparable to those of reported PZT films,[12] which indicate that the PZT phase retains its good quality in the composite film. We also performed magnetization *vs*. field (*M–H*) measurements using a Quantum Design Magnetic Property Measurement System. Figure 3(b) shows the measured *M–H* loop with the magnetic field applied in the direction of the plane. The values of saturation magnetization and coercive field turned out to be about 35 emu/$cm^3$ and 70 Oe, respectively. When we scale saturation magnetization value by the volume fraction of NFO, the normalized value (i.e. 100 emu/$cm^3$) is in good agreement with the reported value for the NFO films (i.e. about 135 emu/$cm^3$).[13]

To determine the $\alpha_E$ values of our PZT–NFO composite film, we performed ME susceptibility measurements using a homemade setup. After poling, the $\alpha_E$ values are deduced from the charge ($\delta Q$) generated from the sample by applying an *ac* magnetic field ($\delta H$) at a frequency of 194 Hz, under the *dc* magnetic bias ($H_B$): $\alpha_E = \delta Q/(\delta H \cdot d \cdot C)$, where *d* and *C* are the thickness and capacitance of the film, respectively. Details of the measurement scheme will be published elsewhere.[14] Figure 4(a) shows the experimental configuration of the transverse components of $\alpha_E$ ($\alpha_{E31}$) schematically, when we measured $\delta Q$//[001] (*i.e.*, along axis '3') while applying both $H_B$ and $dH$ along [100] (*i.e.*, along axis '1'). Figure 4(b) shows a similar schematic diagram for the longitudinal components of $\alpha_E$ ($\alpha_{E33}$), when we applied both the *dc* and *ac* magnetic fields along the [001] direction.

Figure 4(c) and (d) show the magnetic field dependent $\alpha_{E31}$ and $\alpha_{E33}$, respectively. The closed and open symbols indicate the results after poling at +0.8 and –0.8 MV/cm, respectively. We measured $\alpha_{E31}$ and $\alpha_{E33}$ at various poling electric fields, which indicated saturation behaviors at around 0.8 MV/cm. The maximum value of $\alpha_{E31}$ was ~4 mV/cm Oe, while the corresponding value of $\alpha_{E33}$ was ~16 mV/cm Oe. These values are somewhat lower than the reported $\alpha_E$ value (~37 mV/cm Oe) of bulk PZT-NFO particulate composites.[7] The smaller values might be due to the lattice clamping effects by the substrate. Figure 4 also shows sign reversal of both $\alpha_{E31}$ and $\alpha_{E33}$ between the two oppositely poled states. This demonstrated that the ME signals originated from ferroelectric domains with an inversion symmetry breaking relationship.

One of the important issues concerning composite films is the origin of the ME effects. In our film, both $\alpha_{E31}$ and $\alpha_{E33}$ showed sign differences, and the magnitude of $\alpha_{E33}$ was four times that of $\alpha_{E31}$. It has been reported that a single NFO crystal has a positive [001] magnetostriction ($\Delta l/l$) perpendicular to *H*, but a negative value parallel to *H*.[15] Furthermore, the differential change in the magnetostriction with *H*//[001], *i.e.*, $\delta(\Delta l/l)/\delta H$, is four times larger than that with $H \perp [001]$. Therefore, in our composite film geometry, the magnetostriction of the NFO grains along axis '3' cause the lateral length of the PZT matrix to be elongated (shortened) when $H$ is pointing perpendicular (parallel) to [001], resulting in the negative $\alpha_{E31}$ (positive $\alpha_{E33}$) under a finite $H_B$ (>0). Both the observed sign and magnitude changes in the $\alpha_E$ coefficients of our composite film imply that its ME effect should come from the magnetostriction coupled lattice constant change of PZT, namely $\alpha_E \sim \delta P \sim \delta(\Delta l/l)$. [In case of randomly oriented NFO nanoparticles, one would have expected no change in sign and magnitude of $\alpha_E$ coefficients.]

In conclusion, we fabricated nanoparticulate $0.65Pb(Zr_{0.52}Ti_{0.48})O_3$–$0.35NiFe_2O_4$ composite films, which are composed of randomly dispersed $NiFe_2O_4$ nanoparticles in the $Pb(Zr_{0.52}Ti_{0.48})O_3$ matrix. The measured magnetoelectric voltage coefficients of the composite films were comparable to those of bulk particulate composites. Our results also demonstrate that the magnetoelectric effects come from direct stress coupling between the ferroelectric and ferromagnetic grains.

This work was supported by Creative Research Initiatives (Functionally Integrated Oxide Heterostructures) of MOST/KOSEF. The experiments at Pohang Light Source were supported by the MOST and POSTECH. KHK is supported by the Korean Research Foundation (MOEHRD) (R08-2004-000-10228-0).

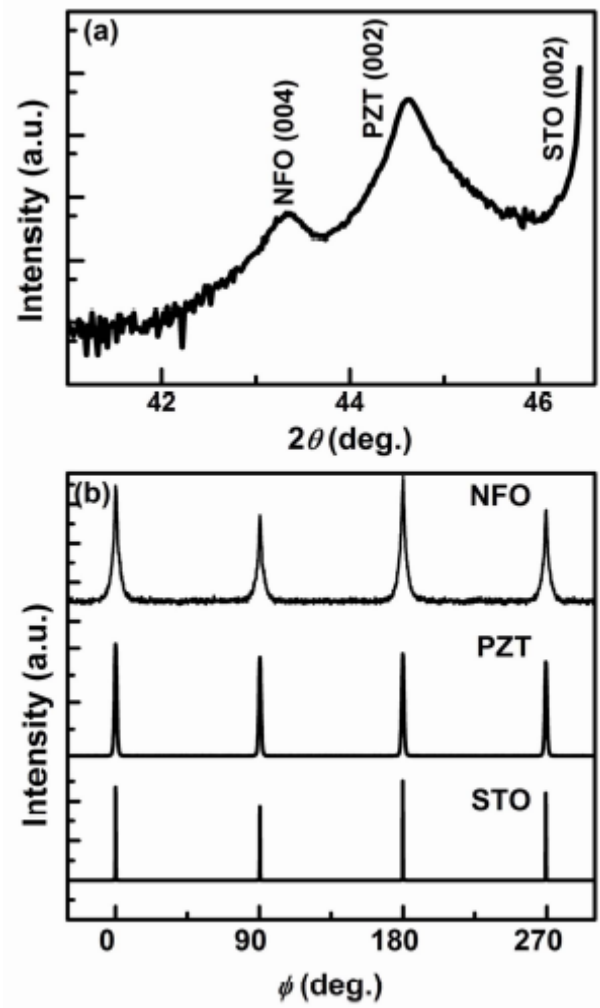

Fig. 1. (a) $\theta$-$2\theta$ XRD scan of a PZT–NFO composite film on Nb:STO substrate. (b) $\phi$-Scans recorded on NFO (404), PZT (202), and Nb:STO (202) reflections.

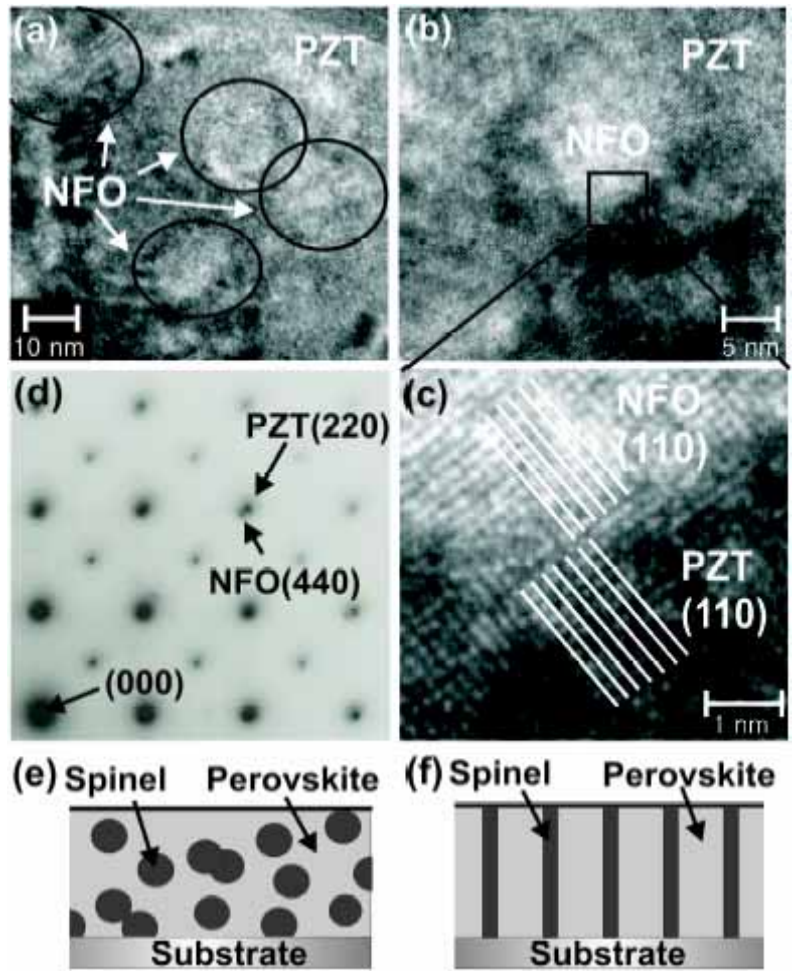

Fig. 2. (a) Low-magnification HRTEM image of the PZT–NFO film. The marked regions indicate the NFO phase. (b) HRTEM image near the PZT–NFO boundary and (c) an enlarged image of the region marked in (b). (d) The selected-area election diffraction (SAED) pattern of the composite film. Schematic representations of a composite film with spinel phase growth involving (e) nanoparticles and (f) self-assembled nanopillars.

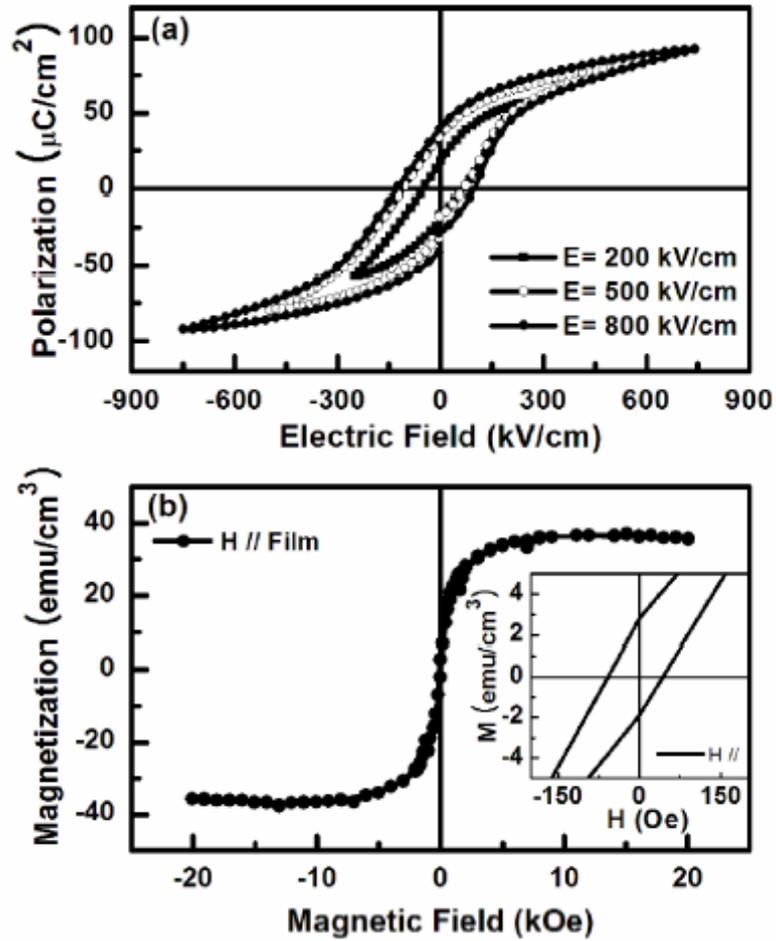

Fig. 3. (a) Polarization hysteresis loops of the composite film measured at various electric fields. (b) The magnetization curve with respect to the magnetic field in the field in-plane direction of the sample.

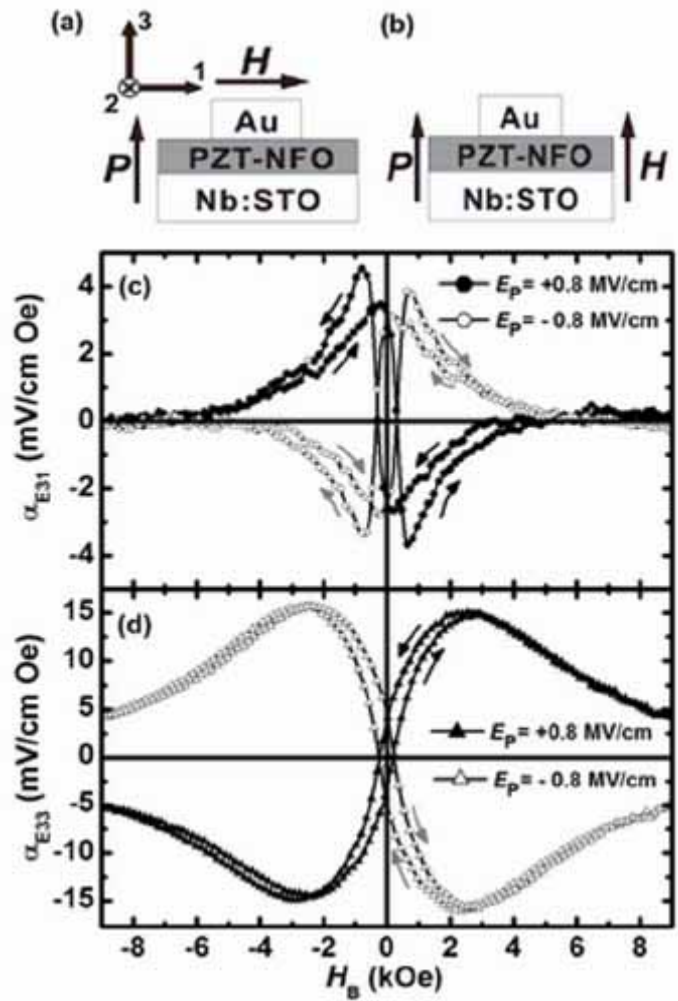

Fig. 4. Schematic diagrams showing the experimental configuration for the (a) transverse ($\alpha_{E31}$) and (b) longitudinal ($\alpha_{E33}$) components of the magnetoelectric coefficient. (c) $\alpha_{E31}$ and (d) $\alpha_{E33}$ measured on the composite film after poling at +0.8 and –0.8 MV/cm.